\documentclass{article}
\usepackage{geometry}                
\usepackage{graphicx}
\usepackage{amssymb}
\usepackage{hyperref}
\usepackage{authblk}

\title{Metropolitan-scale COVID-19 outbreaks: how similar are they?}
\author[1,2]{\small{Samuel Heroy}}
\date{2 April, 2020}                                           
\affil[1]{\footnotesize{The Bartlett Centre for Advanced Spatial Analysis, University College London, London W1 4TJ, UK}}
\affil[2]{\footnotesize{Mathematical Institute, University of Oxford, Oxford OX2 6GG, UK}}
\affil[*]{\footnotesize{Corresponding author: Samuel Heroy, sam.heroy@gmail.com}}

\begin{document}
\maketitle

\begin{abstract}
In this study, we use US county-level COVID-19 case data from January 21-March 25, 2020 to study the exponential behavior of case growth at the metropolitan scale. In particular, we assume that all localized outbreaks are in an early stage (either undergoing exponential growth in the number of cases, or are effectively contained) and compare the explanatory performance of different simple exponential and linear growth models for different metropolitan areas. While we find no relationship between city size and exponential growth rate (directly related to $R0$, which denotes average the number of cases an infected individual infects), we do find that larger cities seem to begin exponential spreading earlier and are thus in a more advanced stage of the pandemic at the time of submission. We also use more recent data to compute prediction errors given our models, and find that in many cities, exponential growth models trained on data before March 26 are poor predictors for case numbers in this more recent period (March 26-30), likely indicating a reduction in the number of new cases facilitated through social distancing. 
\end{abstract}

\section{Introduction}

Why have some cities become hubs - in terms of both cases and deaths - for COVID-19 outbreaks? While the pandemic is no longer in an early stage internationally, it remains of great interest to identify areas where growth is likely to be rapid and (obviously) to determine the overall trajectories of local epidemics. On one hand, we might expect the most connected (internationally and domestically) cities to be affected first and hardest. However, the most rampant case growth for affected areas is more exacerbated by community spread, and it is not altogether clear what city-level factors influence community spread, but it may be expected that city size, city density, and factors involving the built environment may have importance. Epidemics also differ in spreading patterns, with many previous ones having strong dependence on humidity \cite{dalziel2018urbanization}, while the COVID-19 epidemic's dependence is still under debate \cite{wang2020high,luo2020role}. Three of the most affected cities at the point of this writing are New York City, USA, Qom, Iran, and Bergamo, Italy. While all of these metropolitan regions are quite connected domestically/internationally, it's not at all clear that any single city-level factor could identify these specific three cities (against more dense, more populous, and more connected or more humid selections of cities) as amongst the most outbreak-prone. 

In this study, we examine COVID-19 growth rates in different US metropolitan/micropolitan areas, using temporal data that is publicly available from the \href{https://www.nytimes.com/article/coronavirus-county-data-us.html}{New York Times}. The intent of this study is to determine whether there are substantial differences in the spreading patterns of different US cities, or whether local epidemics are only rescaled versions of one another at various time windows with predictable trajectory. On one hand, there are major differences in pandemic-response policies among cities.(many of these policy differences in the US have been imposed through state governments, which are passed to municipalities) in regards to social distancing measures, that both influence flows into/out of the city and also modify contact patterns amongst urban residents. Additionally, another complicating factor lies in geographic differences in testing protocols, and testing availability. On the other hand, there are some natural reasons why scale might be expected to affect the spread of an infectious disease. For instance, studies using cell phone data have found that social network properties (in particular, a person's number of friends) have scaling relations with respect to city size \cite{schlapfer2014scaling,samaniego2020topology}, and studies that simulated pandemics using geotagged mobility networks indicate that larger cities may facilitate more rapid epidemic spreading \cite{tizzoni2015scaling}. Additionally, local/global flows tend to be most concentrated in large cities with centrality in the airline transportation network, which is of critical importance to epidemic spreading \cite{colizza2006role}. Finally, mesoscopic structures in large cities (e.g. conferences, tourist attractions, public transport networks) might be more favorable to disease spread (see \cite{hebert2015complex}). However, there is as yet much debate on the effect of city size and epidemic spreading (especially COVID-19 spreading), with no clear conclusion on the nature or form of this relationship. 

One particular recent study \cite{stier2020covid} - uses US metropolitan infection rates during a period in March to estimate COVID-19 exponential growth rates at the metropolitan level. Interestingly, this study shows a power law scaling relation between city size and infection rate, with larger cities having a more pronounced growth rate in the early phase (they, as we, assume that the epidemic stages are locally early and described well by exponential growth). This has serious implications for both immediate and longterm prognostications of the spread, as the peaks in larger cities will then be higher (as well as potentially sooner) and the overall number of infected people in those cities is also likely to be much higher. However, studies using data from measles (UK) \cite{bjornstad2002dynamics} and influenza (USA) \cite{dalziel2018urbanization} outbreaks have not found the same conclusion, with the former study showing city size has an effect on transmission rate but not on the reproductive ratio and the latter showing that herd immunity may help deter spreading in the largest US cities.

One particular detail of the work in \cite{stier2020covid} is that the modeling framework assumes all cities are in the same phase of the outbreak during the time of study, whereas in reality some cities may still have the outbreak contained during the period, or have otherwise lacked an appropriately representative number of tests. Here, we allow for some variation in the time at which exponential growth begins in metropolitan level outbreaks, in addition to variation in a metropolitan level spreading parameter (as in \cite{stier2020covid}). Additionally, we use two means of fitting - both least squares fitting to logged data (as in \cite{stier2020covid}) and nonlinear least squares fitting - and compare the performance of these models to linear fits. This last issue has potentially large effects with regards to estimating the growth rate (as has been shown in certain urban scaling studies for instance \cite{leitao2016scaling}). We also predict counts for March 26-30, and find that exponential fits based on earlier data tend to overpredict, suggesting social distancing measures may have had considerable effect.

Our analysis provides evidence that - in contrast to the findings of \cite{stier2020covid} but agreement with \cite{bjornstad2002dynamics} (which studies measles outbreaks instead) - larger metros in the US do not experience more rapid exponential growth on average, but are instead later on average in their localized epidemics as of March 25. This finding indicates that while smaller metropolitan areas are currently less affected as of March 25, there is as yet no reason to suspect that the initial phase (exponential growth) of spreading will be less severe in the near future and present. Notably, this analysis only pertains to this initial stage (exponential growth) of localized epidemics, and so our findings do not have any prognosis for understanding when or how much these localized epidemics will slow in growth. Cities can in particular reduce their case growth through social distancing measures in order to `flatten the curve' and escape exponential growth.
\begin{figure}[htbp]
\begin{center}
\includegraphics[width=.97\linewidth]{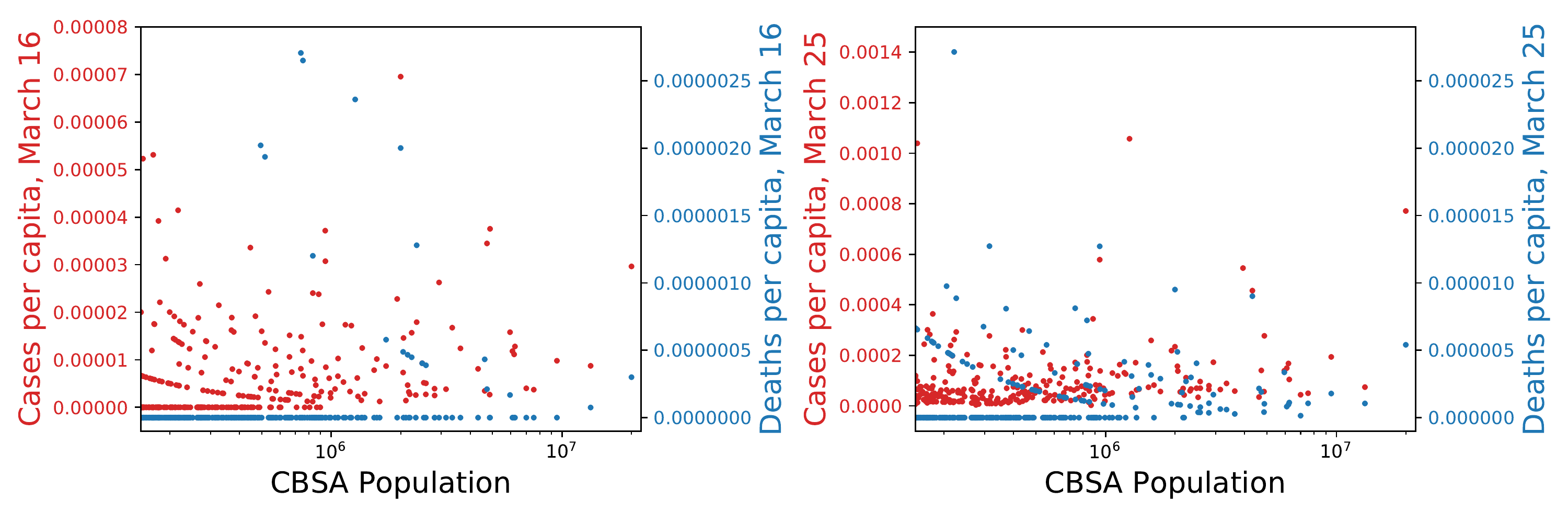}
\caption{\small{Distribution of infections and cases per person against population for two time points.}}
\label{fig:twotimes}
\end{center}
\end{figure}

\section{Data and context}
The data (available \href{https://www.nytimes.com/article/coronavirus-county-data-us.html}{online from the New York Times}) in this study comes a comprehensive data collection effort and includes county-level case/death counts from late January (when the first US cases were identified) through the end of March, 2020. We aggregate infections/deaths from counties (as provided in the raw data) to core-based statistical areas (CBSA) level, geographical area delineations made by the US Office of Management and Budget. These aggregation levels are frequently termed micro/metropolitan statistical areas. We also use 2018 population estimates provided by the US census at the same levels (we suspect these estimates have not changed drastically for most cities over the two year period).

As can be seen in Fig.~\ref{fig:twotimes}, a period spanning just over a week has drastically increased the case/death count for many US CBSAs. Although there have been considerable shifts in the distribution of per capita counts against population, New York City (highest population) has remained very high in cases per capita, while Los Angeles and Chicago (second/third highest) have remained relatively low. Additionally, many cities have experienced very low or zero deaths per capita at both time points, which is to be expected given the lag time between infection and death.
\begin{figure}[htbp]
\begin{center}
\includegraphics[width=.97\linewidth]{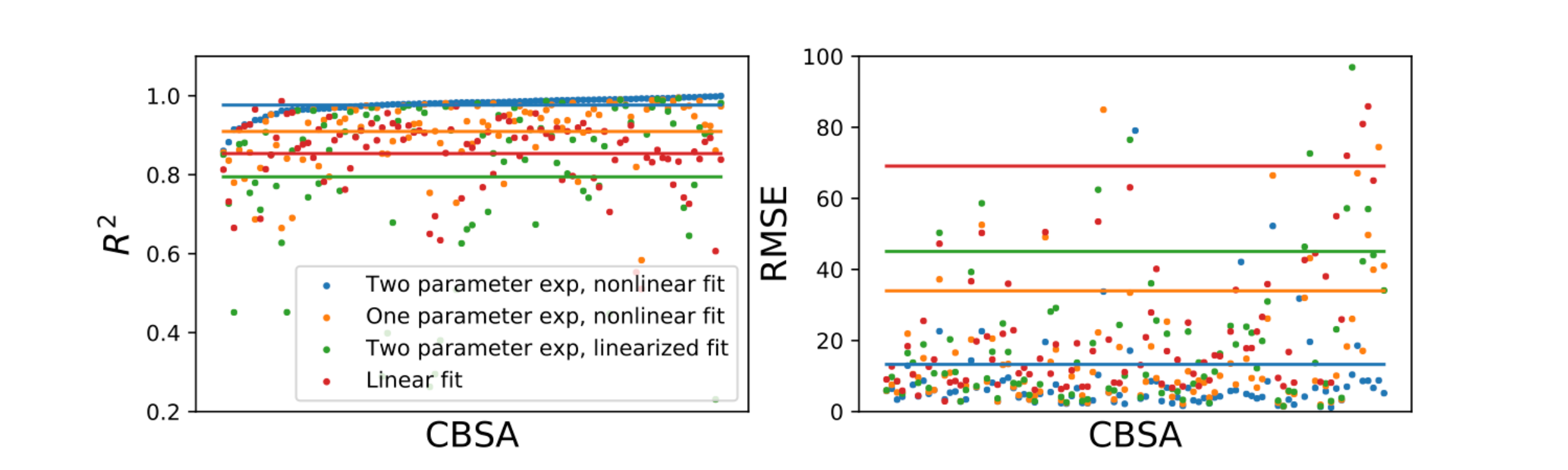}
\caption{Generally, our two parameter exponential fit (through nonlinear least squares) performs best in terms of $R^2$ and RMSE.
}
\label{fig:error}
\end{center}
\end{figure}

\section{Exponential growth estimations}
Here, we estimate city-level growth rates using a simple exponential growth framework, including a second parameter that effectively accounts for the variable stage at which cities are respectively embedded in the pandemic. Specifically, we estimate a two-parameter exponential growth model:
\begin{equation}
I_{t,c}=e^{r_c(t-t_{0,c})}+\epsilon_{t,c},
\label{egrowth}
\end{equation}
where $I_{t,c}$ is the number of infected individuals in CBSA $c$ at time $t-t_0$ days since the outbreak entered a stage of rampant exponential growth in that CBSA, and $r_c$ indicates the CBSA-level exponential growth rate. We note that even though exponential growth is still described by the function for $t_c<t_{0,c}$, the pointwise residuals for these intervals terms will be relatively small ($<1$) and thus likely have little effect (see conclusion for more discussion in this regard). Effectively, this form is equivalent to that given by \cite{stier2020covid} (or any exponential growth model) - taking the log of both sides gives $-t_{0,c}$ as an intercept in the equation above - but we instead fit the model for the entire period in which a CBSA has at least one case (much longer than the five day period used by \cite{stier2020covid}). We use a nonlinear least squares fitting procedure, with initial seeds given by $t_{0,c}$ equal to the day in which the outbreak began in each CBSA and growth rate equal to 0.3. In addition, we use a linear least squares estimation (fitting the logarithm of the left hand side in the exponential growth equation to the right hand side) procedure for comparison. That is, the nonlinear least squares approach minimizes $\sum_t\left(I_{t,c}-e^{r_c(t-t_{0,c})}\right)^2$ whereas the linearized least squares approach minimizes $\sum_t\left(log(I_{t_c})-r_ct-t_{0,c}\right)^2$ for arguments $t_{0,c}$ and $r_c$. We also compare this to a model where this $t_{0,c}$ parameter is not included.  Finally, we also fit for each CBSA a linear model, under the presumption that some CBSAs may not be dominated by community infection - or otherwise may not be in an exponential growth phase - and that a linear model may better describe local growth in these cities. 

We compare the quality of our model fits through the coefficient of determination $R^2$ as well as the root mean square estimation error (RMSE). In general, this comparison would benefit from a likelihood approach that assumes some functional form for the deviations from the fit, but we leave this to other work. Finally, we limit our analyses to CBSAs in which there were at least 50 recorded infections by March 25. While there is no particular reason to choose any specific threshold, estimations based on CBSAs that do not match this criterion are presumed less likely to be accurate as we expect the effect of testing variations to obscure results more in such early stage CBSAs. 

Generally, we find (as anticipated) that the nonlinear least squares fit to the two-parameter exponential growth model is a (marginally) better fit than that of either the 1-parameter approach or the linearized 2-parameter approach, as evidenced by higher values of $R^2$ and lower root mean square errors (Fig.~\ref{fig:error}). In fact, we find this to be true for every CBSA in which we conduct the analysis (mean difference between 2-parameter nonlinear estimation and 2 parameter linearized estimation is $0.018$ for $R^2$ and $-3.25$ infections for the root mean square error). We do find that a linear model is preferable in terms of these criteria in three CBSAs, possibly indicating that epidemic spreading has not reached an exponential stage in these CBSAs (Fig.~\ref{fig:linear}).
\begin{figure}[htbp]
\begin{center}
\includegraphics[width=.85\linewidth]{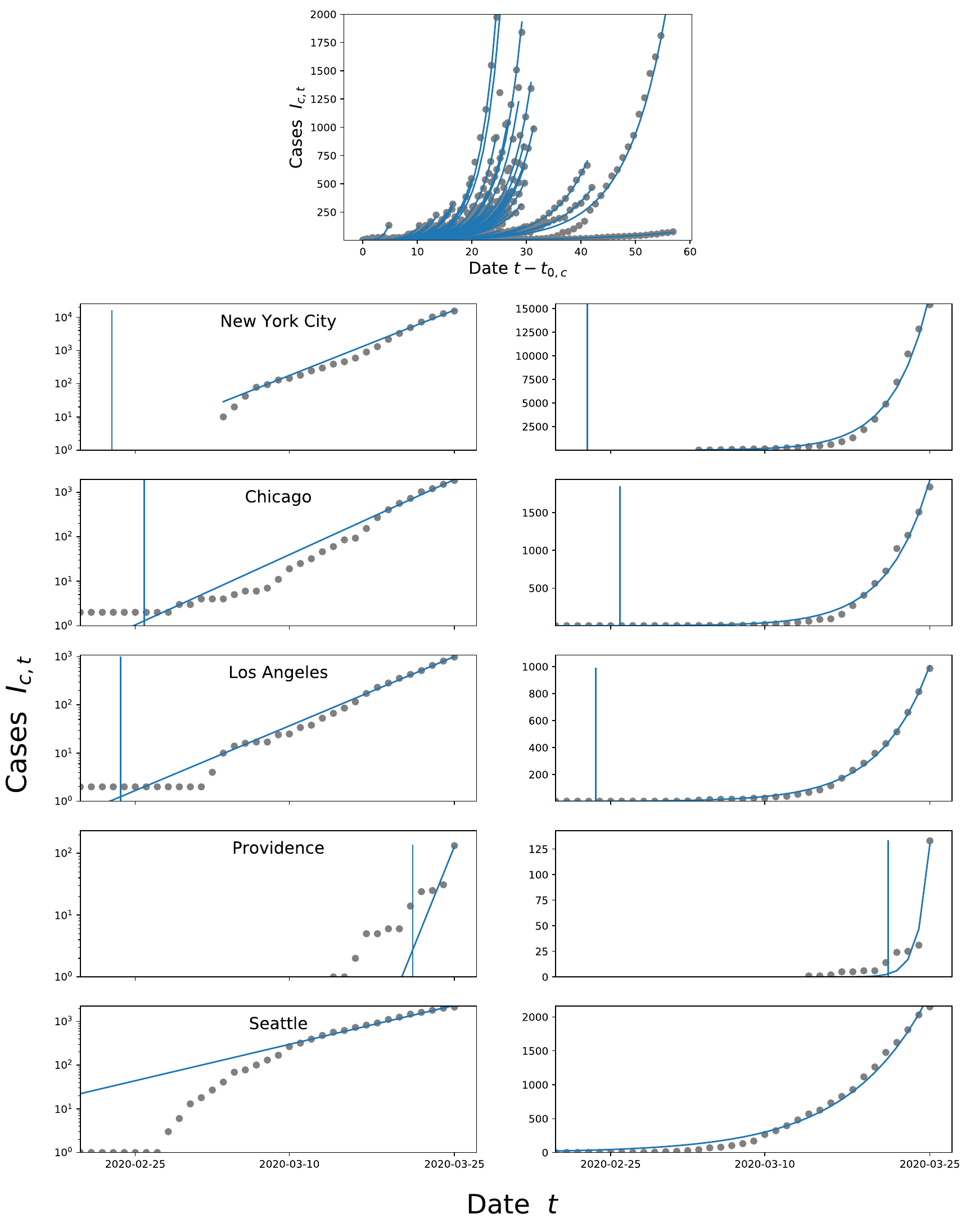}
\caption{\small{\emph{Top:} Number of cases (along with prediction curves) against time rescaled by shifting each CBSA-level trajectory by the estimated value of $t_{0,c}$. \emph{Bottom (8 figures):} Fitted exponential growth models for infection count growth (logarithmically scaled on the left and linearly scaled on the right) in the three largest US CBSAs, as well as one CBSA with an especially late onset of exponential growth and another with an especially early outbreak. Because our fitting procedure minimizes the residual least squares based on raw infection counts, rather than their logarithms, our procedure is better tuned to fit the trajectories on a linear scale. Moreover, this approach (as opposed to the linearized least squares fit approach) perhaps leads to more accuracy for $t>>t_{0,c}$ at the expense of inaccuracy - especially on a log scale - for low $t$ (where residuals will be much smaller on a linear scale for any reasonable exponential fit).
}}
\label{fig:fitted}
\end{center}
\end{figure}

\begin{figure}[htbp]
\begin{center}
\includegraphics[width=.7\linewidth]{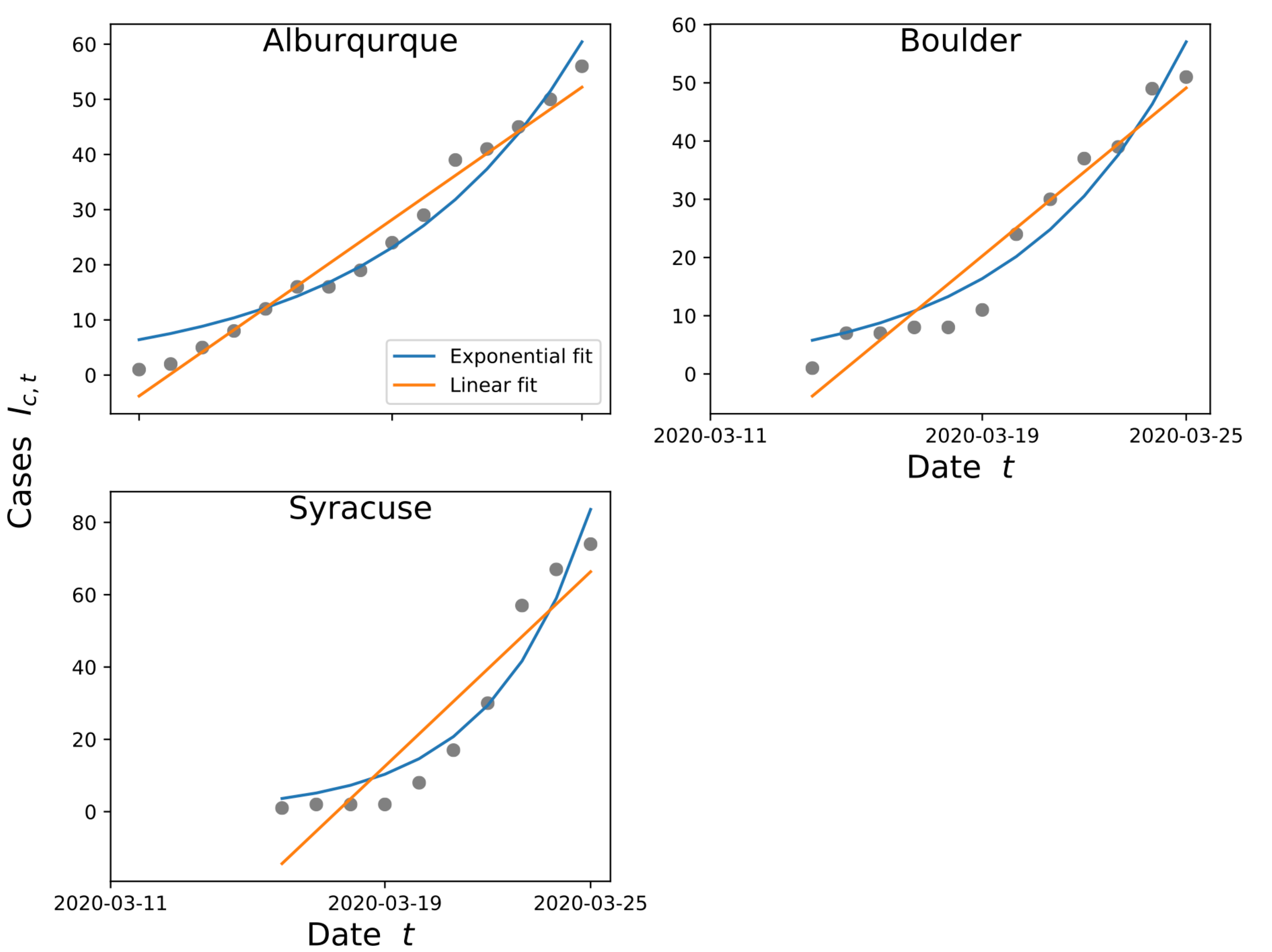}
\caption{\small{In exactly three cases, a linear fit better describes (according to $R^2$, RMSE, or both) CBSA-level trajectories in infection counts.}
}
\label{fig:linear}
\end{center}
\end{figure}

\section{Analysis}
\begin{figure}[htbp]
\begin{center}
\includegraphics[width=.83\linewidth]{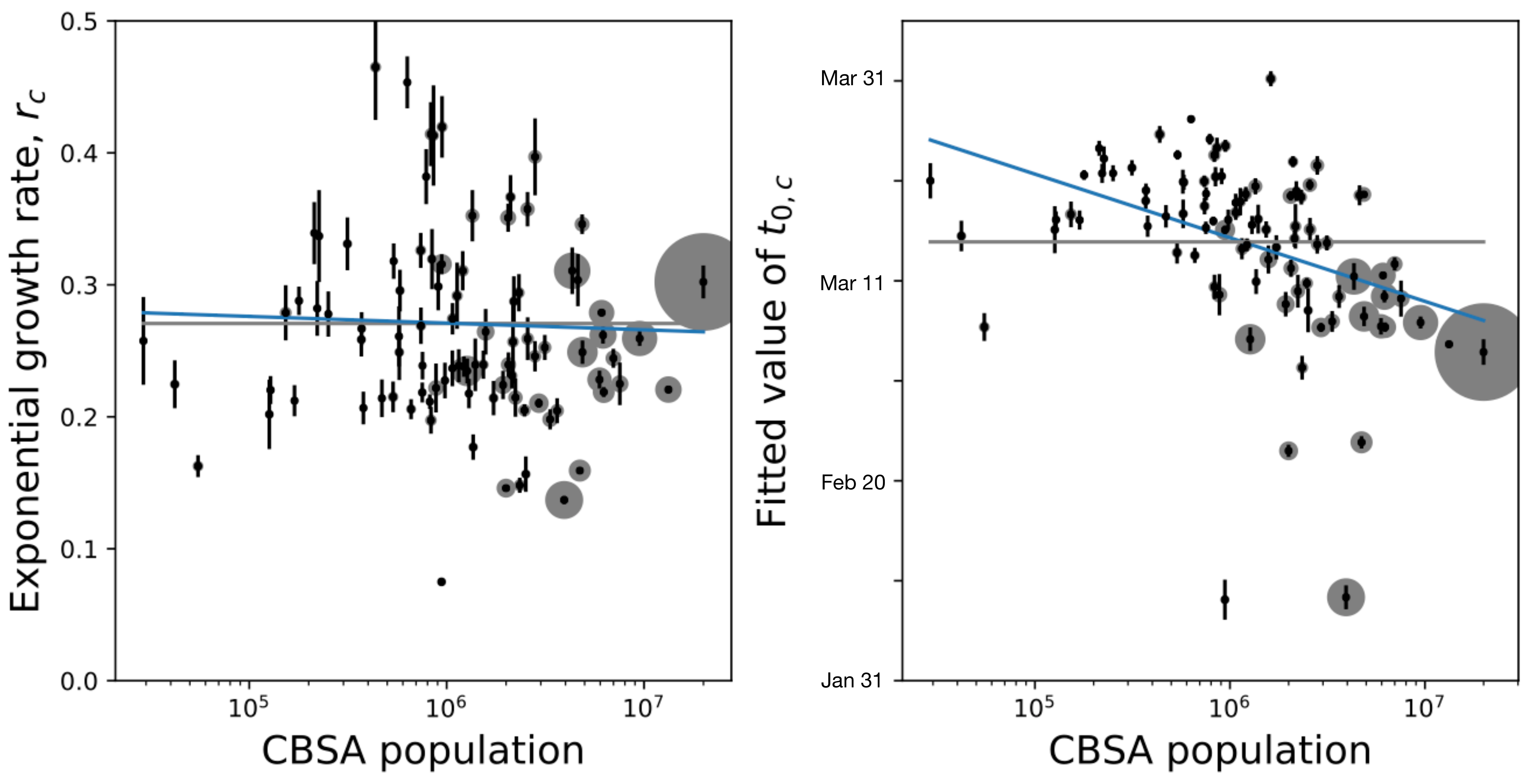}
\caption{\small{Scatter plots of fitted exponential growth rate and $t_{0,c}$ against log($pop_c$) (with sizes proportional to infection counts as of March 25) indicate that larger metropolitan areas do not necessarily have more pronounced exponential growth, but rather that the period of rapid growth began earlier on average in larger cities. Gray lines here represent the average across CBSA-level estimations, while blue lines indicate linear fits ($p>.5$ in the left, $p<10^{-3}$ in the right). Error bars represent one standard deviation above/below the corresponding estimates.}}
\label{fig:pop}
\end{center}
\end{figure}

\emph{Findings with respect to urban population}\\
We examine exponential growth rate - as well as the parameter $t_{0,c}$ - as a function of CBSA population. Importantly, we do not find that the rate has a noticeable dependence on population (Fig.~\ref{fig:pop}-left), but rather the epidemics are locally in later stages in metropolitan areas on average (Fig.~\ref{fig:pop}-right), as evidenced by a decreasing trend in $t_{0,c}$ against population. For example, we note that New York City has become especially concentrated in cases on account of both a low estimated $t_{0,c}$ and above average (though not exceptional) growth rate. Seattle has also become concentrated in infections on account of a particularly early start date - our estimation procedure indicates late January, though this estimation seems possibly erroneous in light of heteroskedasticity in Fig.~\ref{fig:fitted} (alternatively, this may reflect less testing, as Seattle is widely recognized as having had an early major outbreak in the US), even though the growth parameter is estimation is very low ($r\sim .01$). Generally, the trend we observe can possibly be explained by outbreaks taking place earlier in larger cities that have more flows. 

\begin{figure}[htbp]
\begin{center}
\includegraphics[width=.59\linewidth]{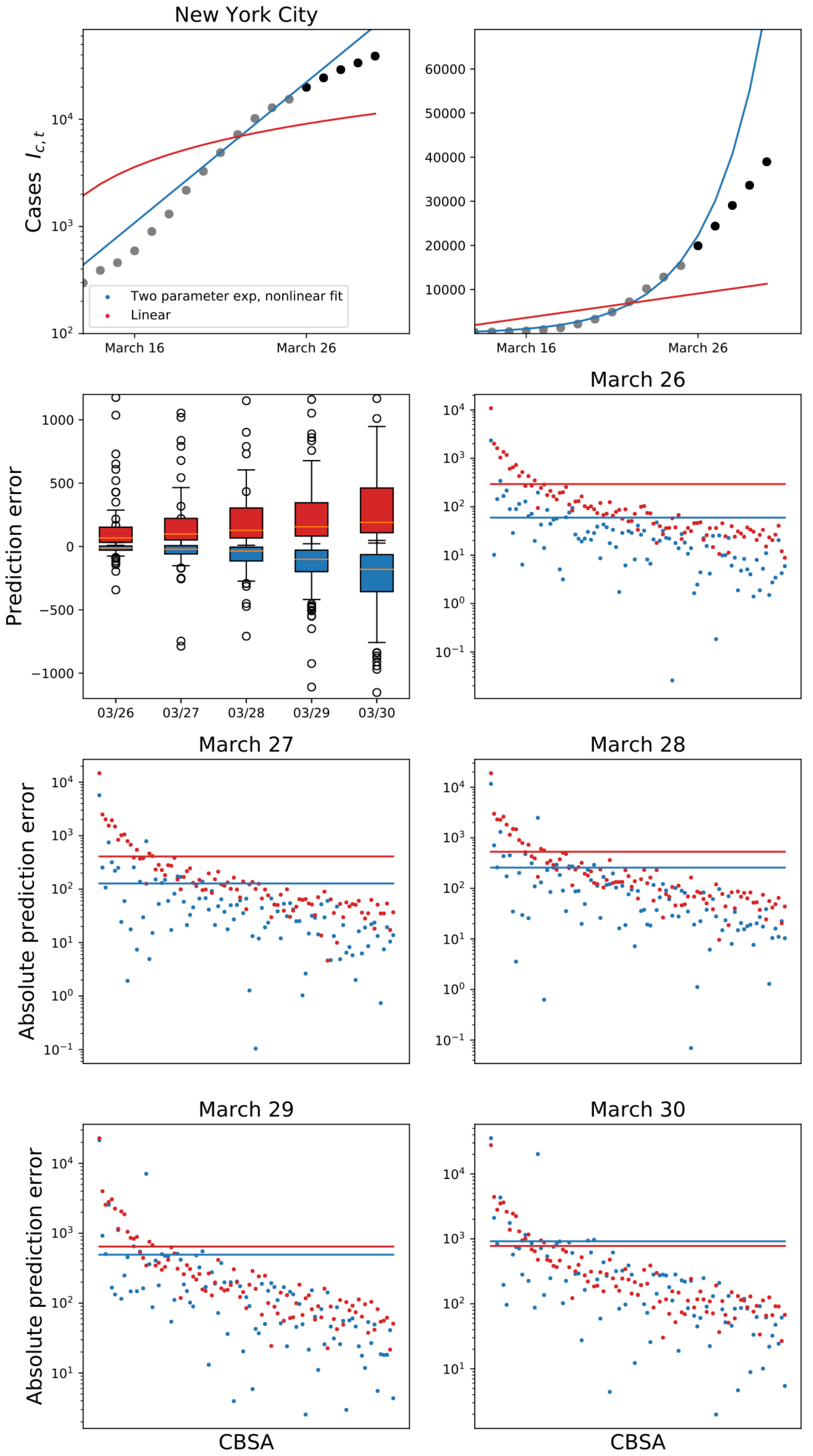}
\caption{\small{\emph{Top row}: Our exponential growth models (trained on data from before March 26) disastrously overpredict the (out of sample) new case counts in March 26-30 in New York City, while linear growth models underpredictsthe same counts, indicating that exponential growth is a decent assumption for the training period, but the nature of the growth fundamentally changed sometime around March 26. \emph{Bottom 6}: Boxplot of prediction error at the CBSA-level for timepoints March 26-30 (as this is a linear scale, a few outliers including NYC fall out of the window), indicating general overprediction by an exponential fit and underprediction by a linear fit. The remaining five plots show absolute prediction error at the CBSA level for the five out-of-sample days. Generally, the exponential model prediction has greater accuracy for most cities early in the out-of-sample window, but the linear fit's accuracy decays more slowly for further out predictions.}}
\label{fig:outsample}
\end{center}

\end{figure}

\emph{Prediction for cases March 26-30}\\
Here we examine how the models have performed in terms of out of sample prediction for new data from March 26-30. That is, we use the models trained from January 21-March 25 to predict the number of cases March 26-30, compute the errors of these predictions given new data, and compare performance (Fig.~\ref{fig:outsample}). We find generally that the exponential growth model tends to overpredict the number of cases in March 26-30, often drastically (e.g. in NYC). While a 2-parameter exponential model (fit via nonlinear least squares) is the optimal model given training data in 92/95 CBSAs that meet our criterion, the linear model outperforms this model in terms of mean square prediction error in 34/92 of these CBSAs. As could be expected, linear models tend to underpredict while exponential models overfit for the vast majority of these predictions - exponentials are nearly always the better fit for one day out predictions, while linear fits are more often better for five day out predictions. We demonstrate this for the important case of New York City in Figure 6 (top) and for all considered cities (bottom). These findings are not at all unexpected, as exponential growth in an epidemic is of course a poor assumption outside of a specific period in which growth is uncontrolled and the number of susceptible individuals unquenched. In particular, these overpredictions might denote effects of increased social distancing measures.

\section{Conclusions and future directions}
Importantly, we find no effect of city size on the exponential growth rate estimated according to our framework and county-level case data from Jan 21-March 25 provided through the New York Times. Instead, we find that while New York City (the highest population CBSA) has only a modestly higher than average growth rate, the exponential growth period  continued for a relatively longer period of time there (especially in comparison to other cities of similar growth rate). This combination of duration and above average growth rate have resulted in New York City being the national concentration hub for Covid-19 infections, but if these findings are to be believed many other US metropolitan areas will soon be grimly facing similar concentrations. However, even if the growth rate in exponential phase is independent of city size, as we find, the duration of this exponential growth period (and other epidemiologically important parameters) will have critical importance to understanding scale in the variation of local epidemic sizes. Our framework assumes that each city is in a purely exponential growth phase, and therefore cannot be used outright to assess the length of this phase. When we make out of sample predictions for March 26-30, exponential fits often lose their appeal as growth has slowed for many CBSAs in this period. 

While we conclude that the major trends highlighted above are robust, certain means of model sophistication might improve this analysis and enhance accuracy at the CBSA-level. Further work might better account for both fitting considerations and to geo-temporal variations in testing protocol/available tests. In regards to fitting considerations, our method for identifying $t_{0,c}$ as the point at which exponential growth became pronounced is rather rudimentary, in that we assume residuals for fits of $f_c(t)$ against $t$ in the interval $t<t_{0,c}$ would have only marginal importance. While this issue did not hamper models' overall high values of $R^2$ and low RMSEs, a change point-based approach might lead to more accurate estimations of the dates at which local epidemics initiated exponential growth (see also \cite{ma2020simple}). Alternatively, with more complete data (especially at a later stage in the localized epidemics), a more appropriate modeling framework (using for instance logistic growth modeling as in \cite{wu2020generalized}) might better capture the underlying spreading nature. Regarding testing considerations, our results are certainly biased on account of the rampant growth in testing availability/capacity over the training period. Indeed, our exponential growth fits generally have a pattern in which residuals ($I_{t,c}-\tilde{I}_{t,c}$) were positive for a short period of time after $t_0$ and negative or zero for some time thereafter, possibly indicating a dearth of tests per number of real cases for the initial period after $t_{0,c}$. A more sophisticated approach might consider this discrepancy and specifically incorporate the number of tests available (such data is made available at \href{https://covidtracking.com}{the COVID Tracking Project}).

\section{Acknowledgements}
The author is supported through the PEAK Urban programme, funded by UK Research and Innovation’s Global Challenge Research Fund, Grant Ref: ES/P011055/1.
\bibliographystyle{siamplain}
\bibliography{references}

\end{document}